\documentclass[Journal,letterpaper]{ascelike-new}
\usepackage{booktabs}
\usepackage{multirow}
\usepackage{makecell}
\usepackage{tabularx}
\usepackage{tabulary}
\usepackage{array}
\usepackage{colortbl}
\usepackage[table]{xcolor}
\usepackage{geometry}
\usepackage{lscape}
\usepackage{multicol}
\usepackage{enumitem}
\usepackage{placeins}
\usepackage{comment}
\usepackage{etoolbox}
\usepackage{float}
\usepackage{tikz}
\usetikzlibrary{shapes.geometric, arrows.meta, positioning}
\tikzstyle{block} = [rectangle, rounded corners, minimum width=7.8cm, minimum height=2.5cm, align=left, draw=black, font=\sffamily\footnotesize, text width=7.8 cm]
\tikzstyle{arrow} = [thick,->,>=stealth]

\usepackage[utf8]{inputenc}
\usepackage[T1]{fontenc}
\usepackage{lmodern}
\usepackage{graphicx}
\usepackage[style=base,figurename=Fig.,labelfont=bf,labelsep=period]{caption}
\usepackage{subcaption}
\usepackage{amsmath}
\usepackage{newtxtext,newtxmath}
\usepackage[colorlinks=true,citecolor=red,linkcolor=black]{hyperref}
\NameTag{Favero, \today}
\AtBeginDocument{\nolinenumbers}

\begin{document}

\title{Calibration and Evaluation of Car-Following Models for Autonomous Shuttles Using a Novel Multi-Criteria Framework}

\author[1]{Renan Favero}
\author[2]{Lily Elefteriadou}

\affil[1]{Department of Civil and Coastal Engineering, 
University of Florida, USA. Email: renanfavero30@hotmail.com}
\affil[2]{Department of Civil and Coastal Engineering, 
University of Florida, USA. Email: elefter@ce.ufl.edu}

\maketitle

\begin{abstract}

Autonomous shuttles (AS) are fully autonomous transit vehicles with operating characteristics distinct from conventional autonomous vehicles (AV). Developing dedicated car-following models for AS is critical to understanding their traffic impacts; however, few studies have calibrated such models with field data. More advanced machine learning (ML) techniques have not yet been applied to AS trajectories, leaving the potential of ML for capturing AS dynamics unexplored and constraining the development of dedicated AS models. Furthermore, there is a lack of a unified framework for systematically evaluating and comparing the performance of car-following models to replicate real trajectories. Existing car-following studies often rely on disparate metrics, which limit reproducibility and performance comparability.

This study addresses these gaps through two main contributions: (1) the calibration of a diverse set of car-following models using real-world AS trajectory data, including eight machine learning algorithms and two physics-based models; and (2) the introduction of a multi-criteria evaluation framework that integrates measures of prediction accuracy, trajectory stability, and statistical similarity, which provides a generalizable methodology for a systematic assessment of car-following models.

Results indicated that the proposed calibrated XGBoost model achieved the best overall performance. Sequential model type, such as LSTM and CNN, captured long-term positional stability but were less responsive to short-term dynamics. LSTM and CNN captured long-term positional stability but were less responsive to short-term dynamics. Traditional models (IDM, ACC) and kernel methods showed lower accuracy and stability than most ML models tested.

\end{abstract}

\section{Practical Applications}

This study provides transportation professionals with calibrated car-following models for AS based on real-world trajectory data and a novel comprehensive framework to evaluate such models. Among several machine learning and traditional models, XGBoost showed the highest accuracy in reproducing AS acceleration and following behavior under mixed-traffic conditions. This model can be applied to simulate AS in urban areas. Using the calibrated XGBoost model, agencies can better predict AS impacts on roadway capacity, operational stability, and safety before deployment.

The new comprehensive evaluation framework developed enables practitioners to assess model accuracy, stability, and responsiveness. By combining real-world data calibration with a systematic evaluation process, this research provides a practical framework for selecting more accurate car-following models, enhancing the fidelity of traffic simulations, and informing planning decisions.

\section{Introduction} 

Car-following models govern longitudinal vehicular motion and form the core of microscopic traffic simulation. Developing accurate car-following models is essential for modeling the impacts of vehicles on traffic networks \cite{Han2021TheReview}, and for public agencies to plan ahead for the needs of an efficient operating road system \cite{Favero2024FactorsPCEs,Yoo2024CombiningVehicles,Favero2017TrafficMaria-RS}. Unlike human drivers, autonomous vehicles (AV) move based on sensor inputs. The type of AV and its underlying technologies can significantly influence car-following behavior \cite{Favero2024FactorsPCEs}.

Autonomous shuttles (AS) are a class of transit AV designed to transport passengers over short to medium distances without a human driver. 
Recent studies have shown that AS exhibit significantly different driving behaviors compared to passenger AV, which necessitates the development of dedicated models \cite{Favero2025}. AS typically operate on fixed routes, at lower speeds, and with conservative acceleration profiles \cite{Favero2025}. These operational characteristics lead to different car-following behaviors compared to both human-driven vehicles (HDVs) and other AV types; thus, using models that are not appropriately calibrated for AS risks oversimplifying or misrepresenting AS traffic impacts.

Despite advancements in car-following models, research addressing AS remains limited \cite{Zhou2024Autoshuttle:Environments}. Existing studies have relied primarily on physics-based approaches, which, although grounded in fundamental principles, often fail to capture the nonlinear dynamics and operational variability characteristic of AS behavior \cite{Zhang2025Car-FollowingReview}. To date, no study has used ML techniques to develop car-following models for AS. ML approaches, when trained using field-collected AS trajectories, provide a data-driven means of representing these complexities.

In this context, developing calibrated car-following models, such as ML-based models, is essential to provide more accurate predictions for replicating AS trajectories \cite{LilyElefteriadou2024SupportTestbed}. Consequently, there is a clear need for well-calibrated AS models to rigorously assess how AS deployment may influence roadway capacity and stability in mixed-traffic environments. Car-following models that generate more stable trajectories are essential to reduce the risk of unrealistic acceleration oscillations.

To evaluate the performance of car-following models in simulating vehicle trajectories, several metrics have been applied \cite{punzo2021}. Studies have shown that these metrics can yield different assessments of model performance \cite{punzo2021}. While some focus on discrete error measures \cite{punzo2021}, others employ alternative evaluation criteria \cite{Zhou2023Data-drivenVehicles}. As a result, there is currently no standardized framework for assessing model performance. Although existing metrics provide valuable but partial insights, there is a clear need to develop a framework that supports multi-criteria assessment, is applicable to any car-following calibration, and enables the comprehensive identification of the most effective model for simulating AS in mixed-traffic conditions. 

This study makes two novel contributions: (1) It provides the first ML–based car-following models calibrated on field data for AS. (2) It developed a multi-criteria evaluation framework that integrates complementary performance metrics into a comprehensive method.
Together, these contributions advance the state of the art in AS modeling and provide actionable guidance for researchers and practitioners. For this study, the following objectives were defined: (1) Develop a set of  ML car-following model based on a field dataset of AS trajectories; (2) Create and apply a multi-criteria evaluation framework integrating accuracy, stability, and trajectory similarity metrics, to provide a comprehensive assessment of car-following models performance; (3) Provide insights for car-following model evaluation, to support researchers and transportation agencies assessing different models.

The paper is organized as follows: a literature review of traditional and ML-based car-following models, followed by the methodology describing data processing, model training, and evaluation. The results section presents performance comparisons and trade-offs, and the paper concludes with key findings and implications.

\section{LITERATURE REVIEW}

Commercially available AV algorithms are generally confidential, as they provide a competitive edge for manufacturers \cite{Milanes2014ModelingData}. Therefore, this literature review focuses on models developed and described in academic research and methods used in previous studies to calibrate car-following models.

This section reviews traditional and ML-based car-following models and commonly used performance evaluation metrics. It concludes by summarizing key findings, research gaps, and the contributions of this study.

\subsection{Traditional Car-Following Models} \label{subsec:traditional_car_following} 

The most commonly used car-following models for AV in the literature are the Intelligent Driver Model (IDM) and Adaptive Cruise Control (ACC) models. The IDM  is widely used to simulate deterministic AV behavior \cite{Favero2025}. However, empirical studies have shown that IDM may not adequately capture responses to speed variations of the preceding vehicle under real-world conditions \cite{Milanes2014ModelingData}.

The ACC model \cite{Milanes2014ModelingData}, incorporated string stability into its design to improve model performance. ACC relies on simplified assumptions, such as speed-error formulations and linear relationships between variables, which limit its ability to accurately reproduce real-world driving behavior. These limitations highlight the need for more adaptable models that can better capture the complexities of real-world driving scenarios.
Although traditional models may not yield the most accurate results, they serve as essential baselines for evaluating more advanced approaches. To date, only one recent study has collected field trajectory data and calibrated car-following models dedicated to AS \cite{Favero2025}. However, its main limitation was reliance solely on physics-based models.

Given the constraints of physics-based models, ML-based methods have emerged as a promising direction for car-following modeling. To date, no study has explored the potential of ML techniques for AS trajectory prediction. 

\subsection{ML for Car-Following Models} \label{subsec:machine_learning_car_following}

ML models have been increasingly adopted for car-following simulation because they can learn directly from data, capture nonlinear relationships, reproduce implicit driving behaviors, and model complex interactions with fewer theoretical assumptions
\cite{Yang2022PhysicalIntersections,Marques2023ShouldData,Zhu2022TransFollower:Transformer}. Despite these advantages, recent reviews highlight key limitations, including reduced physical interpretability, sensitivity to overfitting, dependence on large datasets, error propagation in sequential prediction, and scalability issues in complex scenarios \cite{Zhang2025Car-FollowingReview}. 

The wide variety of ML algorithms allows selecting models tailored to specific problems and datasets. For this reason, this study tested the models: Support Vector Machine (SVM), Random Forest (RF), XGbosst, LightGBM (LGBM), Feedforward Neural Networks (FNN), Convolutional Neural Networks (CNN), Long Short-Term Memory Networks (LSTM), and Transformer, based on their widespread adoption and demonstrated performance in regression tasks.

The SVM models are widely used for regression because they can handle nonlinear relationships with small datasets and require limited parameter tuning \cite{Xue2019RapidData}.

Tree-based and boosting models have shown strong performance in trajectory prediction. RF can achieve high accuracy and robustness, but may struggle to fully reproduce realistic car-following spacing or driver heterogeneity \cite{Shi2021AForest}. 
Studies indicated that XGBoost offers improved computational efficiency and predictive accuracy compared with traditional car-following models, performing well even with relatively small datasets \cite{Zhao2022XGBoost-DNNHighway}. LGBM further improves training speed and scalability, although its application to car-following trajectory prediction remains largely unexplored \cite{Wang2024Risk-QuantificationPropensity}.

Neural network models have also been widely applied to car-following simulation due to their ability to learn complex temporal patterns. FNN have demonstrated improved trajectory prediction when historical information is included and can reproduce both microscopic behavior and macroscopic traffic characteristics \cite{Zhao2025IntegratingManipulator}. CNN can generate smoother speed profiles \cite{Qin2023AAbility,Tang2024ApplicationData}. Recurrent architectures such as LSTM can effectively capture complex traffic phenomena, including platoon dynamics and traffic hysteresis \cite{Lin2022PlatoonModel}. Transformer models have shown superior performance in long trajectory sequences \cite{Zhu2022TransFollower:Transformer}. However, their robustness across diverse and shorter car-following segments remains insufficiently tested. 

Additionally, many hybrid ML approaches rely on datasets that lack full AV trajectories, limiting their applicability to AS contexts \cite{Cheng2025AAnalysis}. 
Because ML approaches differ in their underlying assumptions, data requirements, and generalization capabilities, evaluating a diverse set of models is essential to identify those most suitable for representing AS car-following behavior \cite{Qin2023AAbility}.

\subsection{Evaluating the Performance of Car-following Models} \label{subsec:mop}

The performance of car-following models has been assessed using a wide range of metrics. Most studies rely primarily on point-wise accuracy metrics, such as MAE or RMSE, while relatively few explore mesoscopic patterns such as oscillations and stability \cite{Rowan2025ASimulations}. However, evaluating models based on a single aspect can lead to a misleading assessment of overall performance.

The divergence in evaluation approaches indicates that there is no consensus on a single metric or unified framework capable of assessing multiple aspects of car-following model performance \cite{Zhang2025Car-FollowingReview} , highlighting a significant gap in the current literature.

The main performance metrics identified in the literature are summarized in Table~\ref{tab:evaluation_metrics}. This study selected the most relevant metrics and grouped them into three categories: error prediction, which captures point-wise discrepancies without accounting for temporal error propagation; trajectory stability, which assesses oscillations and instabilities that may lead to unrealistic behavior, shock-wave propagation, and reduced user comfort \cite{Zhou2023Data-drivenVehicles}; and trajectory similarity, which evaluates the alignment of predicted and observed trajectories in terms of timing, overall trend, and cumulative deviations.

\begin{table}
\caption{Summary of Evaluation Metrics Used in Model Assessment}
\label{tab:evaluation_metrics}
\centering
\small
\renewcommand{\arraystretch}{1.3}
\begin{tabular}{p{1.2cm} p{5cm} p{3.5cm} p{5.0cm}}
\hline\hline
\multicolumn{1}{c}{Category Name} &
\multicolumn{1}{c}{Category Description} &
\multicolumn{1}{c}{Metric} &
\multicolumn{1}{c}{Metric Description} \\
\multicolumn{1}{c}{(1)} &
\multicolumn{1}{c}{(2)} &
\multicolumn{1}{c}{(3)} &
\multicolumn{1}{c}{(4)} \\
\hline
\multirow{3}{*}{\centering\shortstack{Error\\Prediction}}
& \multirow{3}{=}{Evaluates point-wise prediction accuracy by comparing observed and simulated values at each time step; does not capture temporal error propagation.}
& Mean Absolute Error (MAE) & Measures the average magnitude of prediction errors \cite{Qu2022AVehicles}. \\
& & Root Mean Squared Error (RMSE) & Penalizes larger errors more heavily \cite{punzo2021}. \\
& & Mean Squared Error (MSE) & Emphasizes large deviations \cite{Qu2022AVehicles}. \\
\hline
\multirow{3}{*}{\centering\shortstack{Trajectory\\Stability}}
& \multirow{3}{=}{Evaluates smoothness, consistency, and stability of predicted trajectories, capturing oscillations, bias, variance, and systematic errors that may affect realism and user comfort.}
& Fast Fourier Transform (FFT) & Detects high-frequency oscillations; lower values indicate smoother predictions \cite{Zhou2023Data-drivenVehicles}. \\
& & Theil's Inequality Coefficient (Theil U) & Decomposes error into bias (B), variance (V), and covariance (C) components \cite{Hourdakis2003PracticalModels,Bai2024EffectsSituations}. \\
& & Coefficient of Variation (CV) & Assesses relative fluctuation consistency. \\
\hline
\multirow{3}{*}{\centering\shortstack{Trajectory\\Similarity}}
& \multirow{3}{=}{Evaluates alignment between predicted and observed trajectories in terms of structure, temporal dynamics, and distributional characteristics, accounting for cumulative and timing errors.}
& Kolmogorov--Smirnov (K--S) Statistic & Captures distributional differences between predicted and actual values \cite{Wang2021InvestigatingTypes}. \\
& & Earth Mover's Distance (EMD) & Measures the effort required to align two distributions \cite{Yang2022FastDistance}. \\
& & Dynamic Time Warping (DTW) & Assesses temporal alignment between real and predicted sequences \cite{Taylor2015MethodApproach}. \\
\hline\hline
\end{tabular}
\normalsize
\end{table}

Relying on a single performance metric may overlook important temporal and dynamic characteristics of car-following behavior. Models with low point-wise errors may still produce unrealistic trajectories due to cumulative errors or instability, while smoother models may fail to capture short-term dynamics. Similarly, good performance on global error metrics does not necessarily guarantee accurate representation of underlying temporal structure.

A recent study introduced a framework combining microscopic error metrics with mesoscopic oscillation measures, highlighting the importance of evaluating compounded errors and intermediate-scale behavior \cite{Davies2024AModels}. However, it did not consider other critical dimensions such as trajectory shape, temporal alignment, and distributional similarity, underscoring the need for a multi-criteria evaluation framework that provides a more comprehensive and practically relevant assessment of car-following model realism and robustness.

\subsection{Literature Review Summary and Research Gaps} \label{subsec:gaps}

Car-following models for AV have been extensively studied, yet the literature reveals clear gaps when applied to AS. Understanding AS behavior is critical, as these vehicles operate differently that HDV and other AV types. 

Previous studies demonstrate the potential of ML to improve car-following model accuracy by capturing complex driving behavior \cite{Zhang2025Car-FollowingReview,Rowan2025ASimulations}. However, advanced ML techniques have not yet been applied to develop AS-specific models, nor is there a standardized framework for comparing models’ ability to reproduce real trajectories, underscoring the need for AS-focused modeling and consistent evaluation methods.

This review identified promising ML algorithms for AS car-following, and relevant performance metrics for model comparison (Table~\ref{tab:evaluation_metrics}). Two key gaps emerge from the literature:
\begin{itemize}
    \item Limited research has focused on modeling or simulating AS trajectories, with no existing studies applying ML techniques to AS field data.
    \item No unified multi-criteria framework exists for systematically comparing car-following models across accuracy, stability, and trajectory realism.
\end{itemize}

\section{METHOD} \label{sec:method}

In this study, we used field-collected AS trajectory data to develop a range of candidate ML models with the potential to accurately replicate the AS car-following process and compare them with calibrated traditional car-following models. The next subsections describe each step.

\subsection{Dataset Description and Cleaning}

The dataset used in this study consists of real-world AS trajectories collected in Lake Nona, a planned community in Orlando, Florida, and is detailed in \cite{Favero2025}. Data were obtained from the AS’ on-board GPS sensors and from a human-driven vehicle (HDV) traveling directly ahead of the AS. Position data were recorded and used to derive speed, acceleration, and positional changes. The dataset included more than four days of operations along an urban road segment. Preprocessing included selecting only trajectory segments where the leader vehicle remained within the detection range of the AS sensors, ensuring valid car-following interactions.

Since outliers can significantly affect ML training, unrealistic data points were removed following procedures from previous research \cite{Favero2025}. A Kalman filter was then applied to smooth trajectories for both leader and follower vehicles, mitigating inherent GPS noise in line with recommended practice \cite{Punzo2005NonstationaryData}. New trajectory segments were defined whenever time gaps between consecutive observations exceeded two seconds, preventing distortions from extended missing data or accumulated error.

After data cleaning, approximately 4,000 seconds of valid trajectory data were obtained. The AS and HDV trajectories were synchronized at each time step, and kinematic equations were applied to compute car-following variables such as speed difference (\(\Delta v\)), spacing (\(\Delta s\)), previous time-step follower acceleration (\(a_{f,t-1}\)), and previous time-step follower speed (\(v_{f,t-1}\)). These variables capture essential aspects of car-following dynamics and served as inputs to the modeling process.  

This dataset is unique because AS field trajectory data are rarely available in the public domain, and to our knowledge, this represents one of the first studies focused on ML modeling applied to AS trajectories.

 In the filtering process, the state transition matrix was dynamically updated to reflect kinematic relationships, enabling smooth and accurate estimation despite irregular sampling. Only the position feature was directly observed, while speed and acceleration were inferred through filtering. To account for process uncertainties and measurement noise, the process noise covariance matrix (\( Q \)) and measurement noise covariance matrix (\( R \)) were tuned as follows:

\[
Q =
\begin{bmatrix} 
0.1 & 0 & 0 \\ 
0 & 0.01 & 0 \\ 
0 & 0 & 0.001 
\end{bmatrix}, \quad
R =
\begin{bmatrix} 
0.5 
\end{bmatrix}
\]

The values in \( Q \) were chosen to permit minor variations in position, moderate adaptability in velocity, and minimal fluctuations in acceleration, thus promoting numerical stability without over-smoothing the motion. The value \( R = 0.5 \) was selected to balance sensitivity to GPS updates and robustness against measurement noise. 

The final dataset was split into 75\% training and 25\% testing, with the training portion further divided (80\% training, 20\% validation) to support model development.

\subsection{Models Development and Training Process}

The training process included the input features \(\ delta(v)\),  \(\ delta(s)\), previous time-step ($t-1$) follower acceleration \(a_{ (t-1)}\), and previous follower speed \(v_{f (t-1)}\) to predict the AS acceleration in \textit{t}. The same features were used for validation and testing. 

The models were trained to predict the variable follower acceleration \(a_{ (t)}\) for two reasons: (1) to align with the typical formulation of traditional models like IDM and ACC; and (2) some machine learning models (e.g., RF and XGBoost) are inherently designed to learn mappings between input features and a target variable, making it challenging to minimize errors in non-direct outputs such as spacing or speed variables. The MSE of acceleration was used as the loss function.

The calibration of the traditional car-following models, ACC and IDM, was performed using a Genetic Algorithm (GA) to identify the parameter set that minimized the MSE between the model-predicted and observed accelerations. The detailed calibration procedure for both models is described in a previous study \cite{Favero2025}.

For the ML-based car-following models, the hyperparameter optimization was performed using the \textit{Optuna} library, selected for its efficient exploration and pruning of large parameter spaces \cite{Akiba2019Optuna:Framework}. Optuna’s Tree-structured Parzen Estimator (TPE) sampler was used with 500 trials per model to explore the parameter space, and an early-pruning strategy discarded unpromising trials. The selected values represent near-optimal solutions within the search space, and were used to retrain each model on the full training dataset.

The validation was conducted using 5-fold cross-validation, employing the \textit{TimeSeriesSplit} function from the \textit{scikit-learn} library to preserve the temporal order of the trajectories. Predictions on the scaled training data were then inverse-transformed back to their original scale for evaluation when a scaler was applied. The neural network-based models (FNN, CNN, LSTM, and Transformer) were implemented using the \textit{Keras} library. Architectural details (e.g., number of layers, units, dropout) are described in the following subsections, along with the optimized hyperparameters obtained from the search.

 The predicted acceleration was numerically applied to update the follower’s speed and position using standard kinematic equations. In this numerical simulation, the leader’s trajectory was obtained directly from the observed field data and treated as an external input to the simulation. The follower’s initial conditions—position, speed, and acceleration—were matched to the first observed values in each trajectory segment to ensure consistency between simulated and real-world conditions.

Once initialized, the simulation advanced recursively, where each model predicted the follower’s acceleration based on the car-following features ($\Delta v$, $\Delta s$, $a_{f,t-1}$, $v_{f,t-1}$). To validate the model outputs, the simulated trajectories were compared against observed trajectories across the entire horizon, with performance evaluated using the multi-criteria framework described in Section~\ref{sec:model_eval}. This ensured that models were assessed not only on point-wise prediction accuracy but also on their ability to reproduce realistic trajectory dynamics. 

The assumptions applied in the training and testing considered only trips longer than 60 seconds to ensure stability in parameter estimation. The acceleration predicted was bounded within physically plausible results, which prevents unrealistic values but may exclude extreme behaviors. The training step size was fixed at 1 second, to balance accuracy and computational cost. While these assumptions may influence the absolute performance of individual models, they are applied consistently across all models and scales, meaning the comparative insights and trade-offs reported in this study remain valid

\subsection{Optimized Model Configurations} 

The training, validation, testing, and hyperparameter optimization (using the \textit{Optuna} library) were applied consistently to all models tested. For each algorithm, the best-performing architecture and hyperparameter configuration were identified through this process. The following subsections describe the architectures and parameter settings that yielded the best results for each model. All the calibrated models, code used, and optimal set of hyperparameters tuned in this research are available in \url{https://github.com/renanfavero30/AS_car_following.git}.

The SVM model was developed using the \textit{Support Vector Regression}, function from the \textit{sklearn} library.
The hyperparameter optimization resulted in the selection of the RBF kernel and a high C value, indicating that the dataset exhibits non-linear relationships, which necessitate a more complex model. 
The SVM tuning resulted in 2069 support vectors, indicating that the SVM model relies on a significant portion of the training data to define the hyperplane in the feature space and has a relatively high dependency on the training samples. 

The RF model was trained using \textit{RandomForestRegressor} from the \textit{scikit-learn} library. The hyperparameter optimization resulted in a large number of trees (776) and a high maximum depth (146), indicating that the dataset is complex and requires a detailed model. The moderate values for minimum samples split (5) and minimum samples per leaf (3) prevented excessive model complexity and avoided overfitting.

The XGBoost model was developed using the \textit{XGBRegressor} function from \textit{xgboost} python library. 
The high L1 regularization (0.946) suggests that some features are pruned automatically to improve the model's generalization.  A high learning rate (0.477) and moderate boosting rounds (125) mean that the model is converging quickly without excessive trees.  

The LGBM hyperparameter tuning yielded a moderate learning rate (0.1), aiming for stable convergence, while the minimum data per leaf (5) prevented excessive model complexity and overfitting. The full feature utilization indicates that all input variables contribute meaningfully to the predictions, while the bagging method introduced controlled randomness, enhancing generalization.

The FNN model optimization resulted in 4 fully connected layers and 41 neurons per layer. A dropout rate of 0.0723 indicates that regularization was applied while avoiding excessive dropout. The Adam optimizer, known for its robust performance in handling non-convex problems, was chosen for its ability to consider different data distributions. A learning rate of 0.0127 enabled faster convergence, while a batch size of 32 and 20 training epochs ensured an optimal trade-off between computational efficiency and learning capacity.

The CNN model was developed using a convolutional architecture adapted for time-series forecasting, with 2 Conv1D layers and 2 dense layers with 32 neurons each, refining the extracted features. A batch size of 32 and 10 epochs indicates a training strategy that prioritizes efficiency.

The LSTM model optimization resulted in a sequence length of 5 time-steps, suggesting that short-term dependencies are more relevant for prediction than long-term dependencies. The dense layers with 64 and 32 neurons refine the extracted temporal features, ensuring a smooth transition from sequence processing to final predictions.
The model employs the \textit{ReLU} activation function, which enhances learning efficiency by mitigating gradient vanishing issues. A batch size of 32 and 15 training epochs suggests a balance between computational efficiency and generalization, ensuring the model learns effectively without overfitting.

The Transformer model architecture combined a 1D convolutional layer (Conv1D) for capturing localized temporal patterns with a Transformer encoder block that applies multi-head self-attention to extract global dependencies. Within the Transformer block, feed-forward layers enhance feature representation, while layer normalization and residual connections ensure stability.

\subsection{Development of New Multi-Criteria Evaluation Framework} \label{sec:model_eval}

The evaluation of model performance across multiple metrics poses a significant challenge, as no single model consistently outperforms others under all criteria. Figure~\ref{fig:metric_comparison} demonstrates that model performance varies substantially depending on the selected metric.. To address the need for a consistent evaluation of several models under multiple criteria, and given the absence of a comprehensive assessment method, this section proposes a unified and objective framework that systematically quantifies and compares model performance using a multi-criteria approach. The approach is based on computing a Z-score to represent model performance, where lower values indicate better performance. The framework integrates multiple complementary metrics, as listed in Table~\ref{tab:evaluation_metrics}. The framework proposed and applied is described step-by-step as follows:

\begin{enumerate}
    \item Calculate each model's prediction error, trajectory stability, and trajectory similarity metrics (described in Table \ref{tab:evaluation_metrics}).
    \item Normalize all metrics using Z-score transformation:
    \begin{equation}
        Z_i = \frac{X_i - \mu}{\sigma}
    \end{equation}
    where:
    \begin{itemize}
    \footnotesize
    \renewcommand{\labelitemi}{}
     \footnotesize 
        \item  \( X_i \): is the observed metric, 
        \item \( \mu \): is the mean of the metric calculated, and 
        \item \( \sigma \): is the standard deviation of each model's predictions.
    \end{itemize}
    
    \item Adjust the scales so that lower values consistently indicate better performance, and stability metrics were expressed as relative differences from the original follower trajectory to ensure comparability.
    \item Compute the acceleration, speed, and position scores by averaging their normalized metric components. The general Equation~\ref{eq:score} was applied to calculate scores for acceleration, speed, and position predicted. 
    
    \begin{linenomath}
    \begin{equation}
    \label{eq:score}
    Z~Score_{\text{j}} = \sum_{i \in \mathcal{E}} \frac{Z_i}{|m|}
    \end{equation}
    \end{linenomath}
where:  
\begin{itemize}
    \footnotesize
    \renewcommand{\labelitemi}{}
    \item \( Score_{j} \): score for each model under aspect \( j \) (error, stability, or similarity),  
    \item \( \mathcal{E} \): set of metrics associated with aspect \( j \) (e.g., for error: \( \mathcal{E} = \{\text{MAE}, \text{RMSE}, \text{MSE}\} \)),  
    \item \( Z_i \): normalized value of each metric \( i \) in \( \mathcal{E} \),  
    \item \( m \): total number of metrics in \( \mathcal{E} \).  
\end{itemize}  
\end{enumerate}

\section{RESULTS}
 
 The following section presents the results related to prediction distributions, performance metrics, and the overall score for each model.

\subsection{Candidate Model Comparison Using Selected Performance Metrics}

\begin{figure*}[htb]
    \centering
    \captionsetup{skip=2pt} 
    \hspace*{-2cm}
    \begin{subfigure}{\textwidth}
        \centering
        \includegraphics[width=1.15\textwidth]{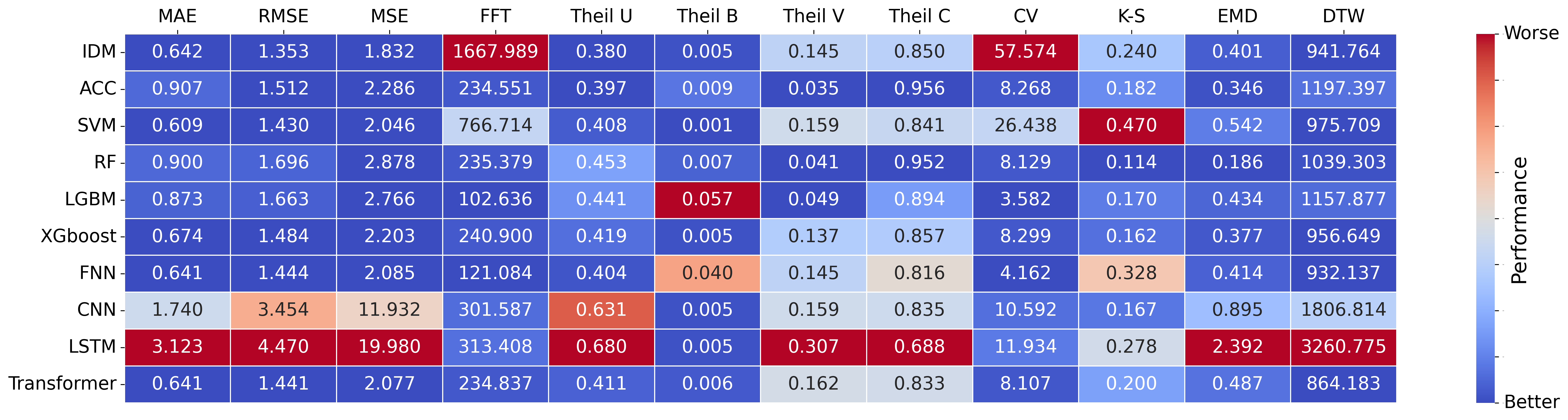}
        \caption{Metric calculated considering the acceleration predicted.}
        \label{fig:error_score}
    \end{subfigure}
    
    \vspace{3pt} 
    \hspace*{-2cm}
    \begin{subfigure}{\textwidth}
        \centering
        \includegraphics[width=1.15\textwidth]{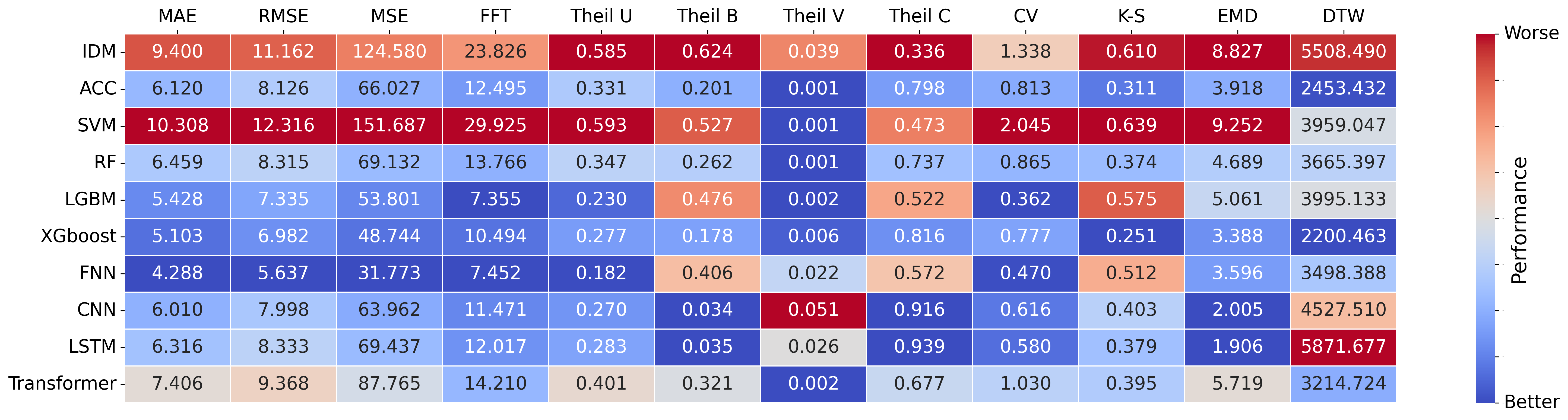}
        \caption{Metric calculated considering the speed predicted.}
        \label{fig:stability_score}
    \end{subfigure}
    
    \vspace{3pt} 
    \hspace*{-2cm}
    \begin{subfigure}{\textwidth}
        \centering
        \includegraphics[width=1.15\textwidth]{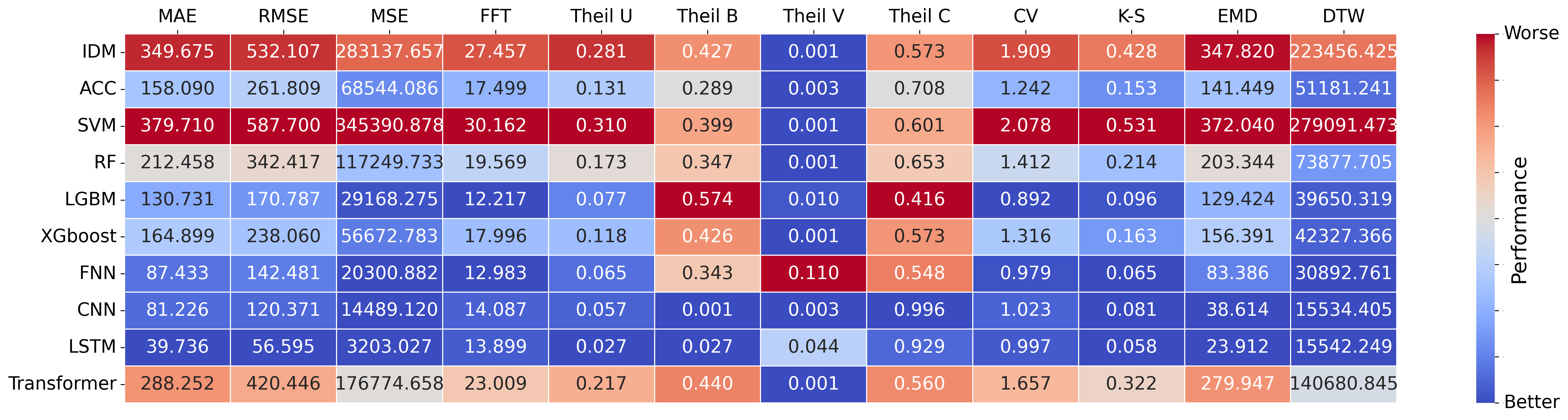}
        \caption{Metric calculated considering the position predicted.}
        \label{fig:similarity_score}
    \end{subfigure}
    
    \caption{Metrics (non-normalized) calculated to obtain the error, trajectory stability, and trajectory similarity scores.}
    \label{fig:metric_comparison}
\end{figure*}

This section presents the model comparison results based on each evaluation metric defined in Table~\ref{tab:evaluation_metrics}. Figure~\ref{fig:metric_comparison} displays each model’s performance considering these metrics using a heatmap. In this figure, blue indicates better performance (i.e., lower error or higher correlation), while red denotes poorer performance. Most numerical metrics are interpreted as "lower is better," except for Theil C, where higher values indicate better trend alignment. This distinction was appropriately incorporated into the color scaling to obtain a consistent representation in Figure~\ref{fig:metric_comparison}.

Overall, the results indicate that models performing well under error-based metrics often underperform in stability, temporal consistency, or trajectory realism. For instance, IDM achieved low MAE and RMSE for acceleration but exhibited unstable and oscillatory behavior (high FFT and CV), indicating overreaction to changes and poor dynamic representation. In contrast, the ACC model showed slightly higher errors but lower Theil U across acceleration, speed, and position, suggesting a better capture of average AS dynamics. These findings confirm that no single model consistently outperforms others across all metrics, underscoring the need for multi-dimensional evaluation of AS car-following performance.

Although the metric-based analysis (Figure~\ref{fig:metric_comparison}) provides valuable insights, selecting the most suitable AS car-following model remains challenging due to the diversity of performance dimensions. To address this issue, the following sections describe the results of the proposed multi-criteria framework, which links observed outcomes to the underlying architectures and learning mechanisms of each model.

\subsection{Multi-criterial Evaluation Results}

This method produces an overall Z-score to identify the model with the best average performance, while also revealing that individual models exhibit strengths in specific aspects and weaknesses in others, highlighting trade-offs across model architectures and performance dimensions.

\begin{figure}[htbp]
  \centering
  \captionsetup{skip=2pt}

  \begin{subfigure}{0.49\textwidth}
    \centering
    \includegraphics[width=\linewidth]{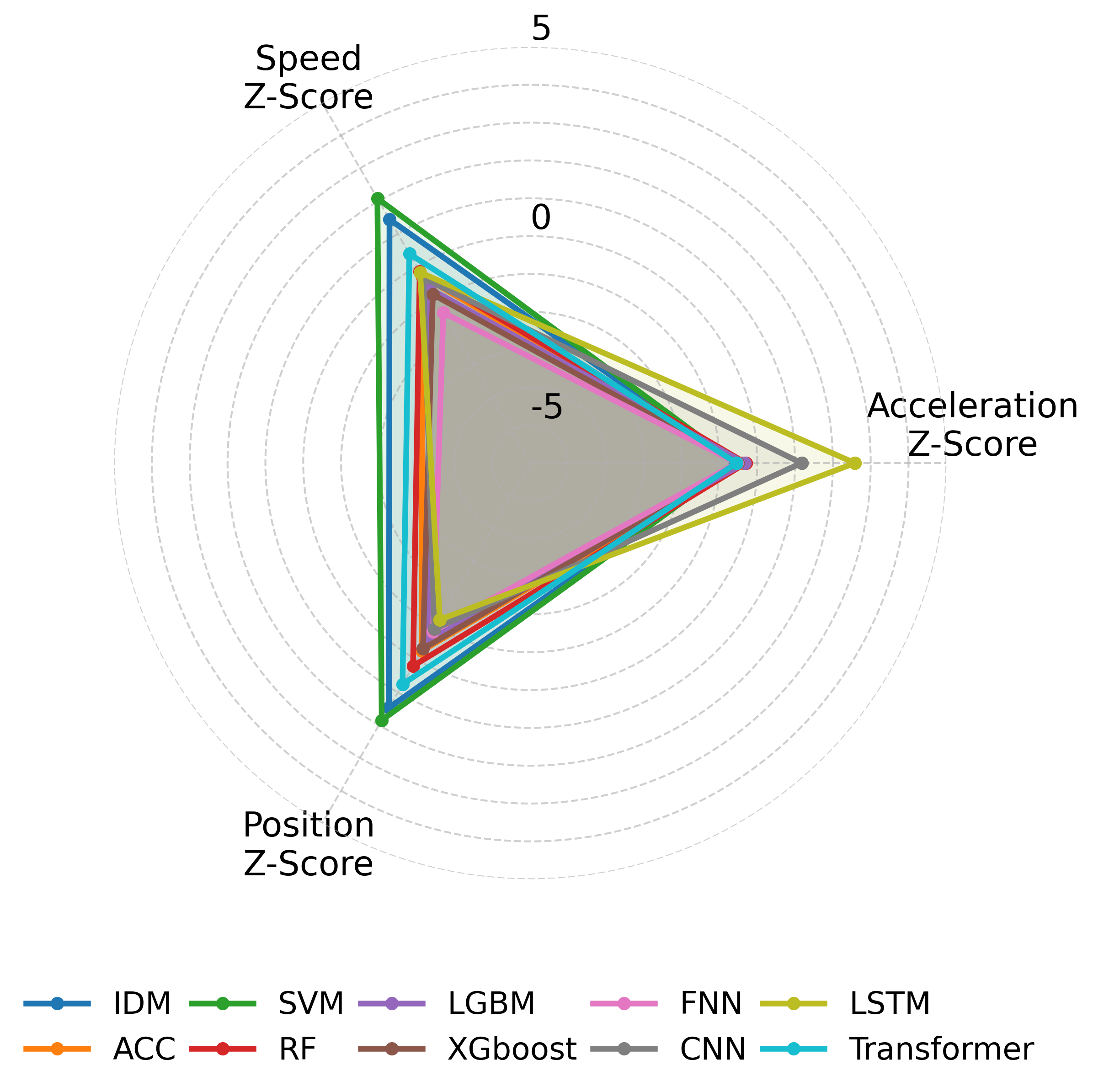}
    \caption{Z-Score for Error Prediction}
    \label{fig:error_score}
  \end{subfigure}
  \hfill
  \begin{subfigure}{0.49\textwidth}
    \centering
    \includegraphics[width=\linewidth]{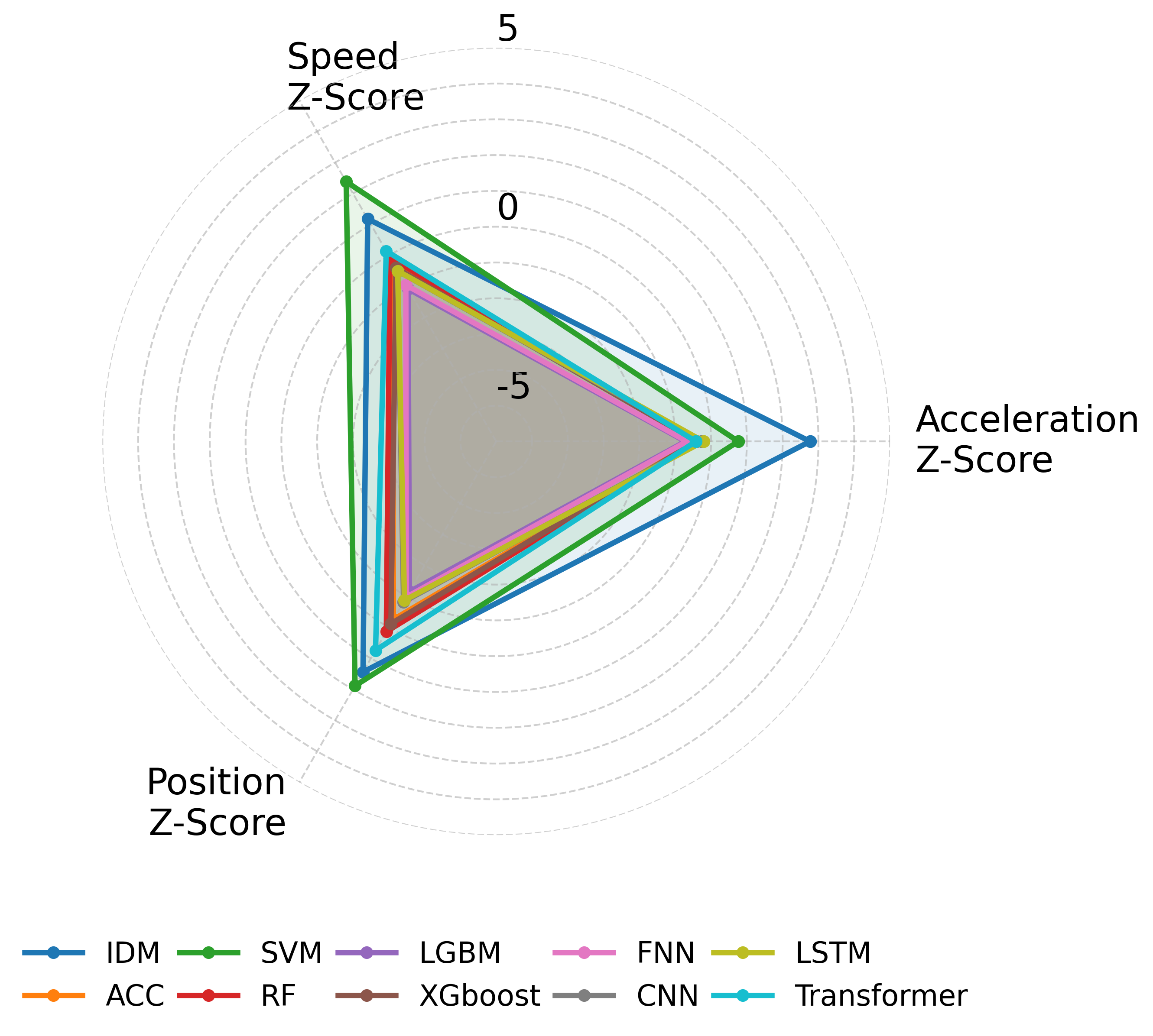}
    \caption{Z-Score for Trajectory Stability}
    \label{fig:stability_score}
  \end{subfigure}

  \medskip

  \begin{subfigure}{0.5\textwidth}
    \centering
    \includegraphics[width=\linewidth]{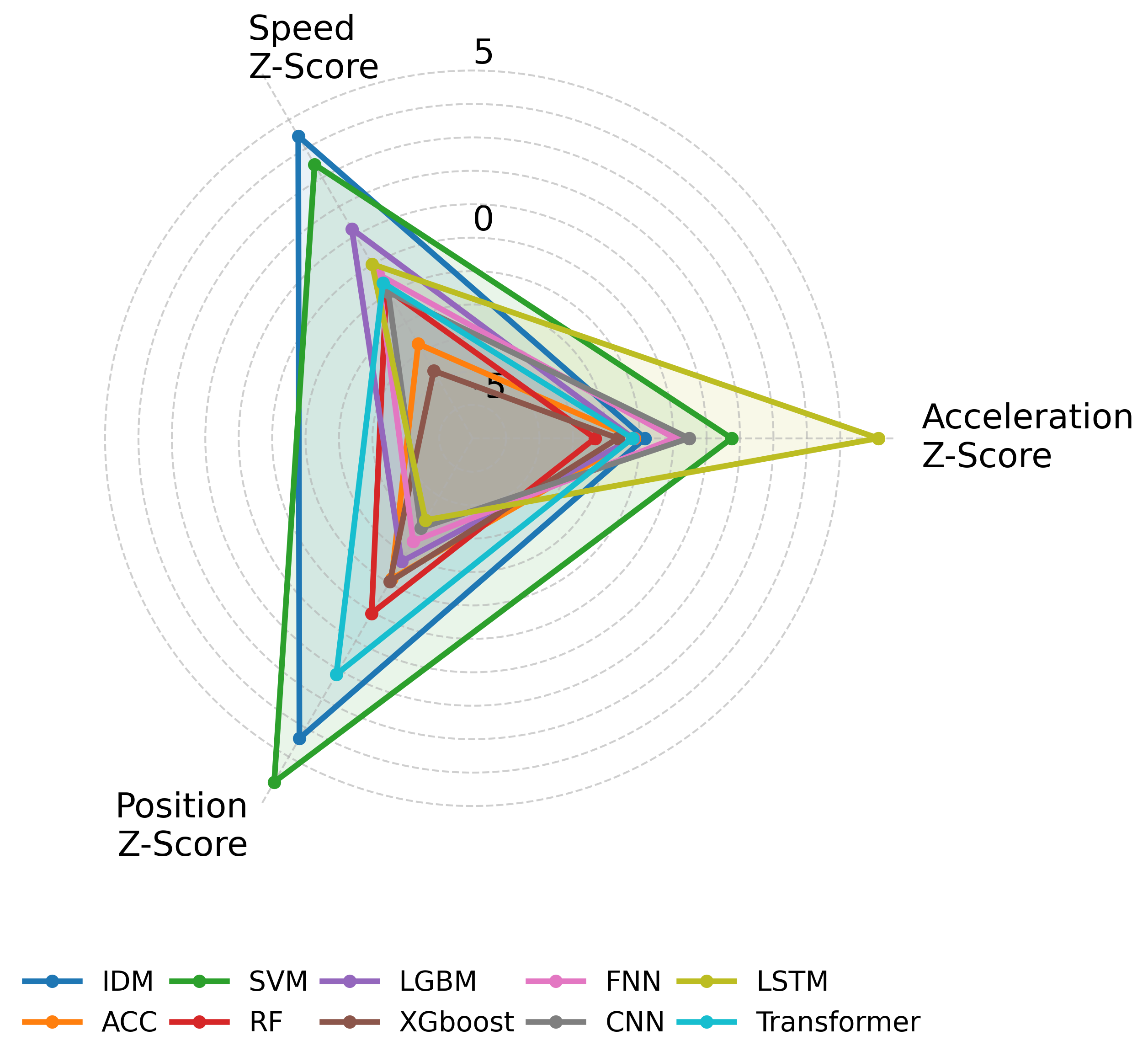}
    \caption{Z-Score for Trajectory Similarity}
    \label{fig:similarity_score}
  \end{subfigure}

  \caption{Aggregated Z-scores across evaluation dimensions.}
  \label{fig:z-score}
\end{figure}

Figure~\ref{fig:z-score} shows in radar charts the models' performance estimated with the multi-criterial method applied. The results from the \textbf{error scores} category (Figure~\ref{fig:error_score}) indicated that the models FNN, XGBoost, and LGBM achieved the lowest error, indicating high predictive accuracy. \textbf{Stability scores} results (Figure~\ref{fig:stability_score}) indicated that FNN, XGBoost, and LGBM are the most robust, with consistently low Z-scores, indicating smooth and stable outputs. LSTM is stable in position but unstable in acceleration and speed, while IDM and SVM exhibit the highest instability. \textbf{Trajectory similarity scores }(Figure~\ref{fig:similarity_score}) indicate that Transformer, FNN, and LGBM best reproduce trajectory shape and distributions. While SVM, IDM, and LSTM perform worst—particularly in position—showing substantial deviation from real trajectories.

\begin{figure}[H]
    \centering
    \captionsetup{skip=2pt}
    \includegraphics[width=0.8\textwidth]{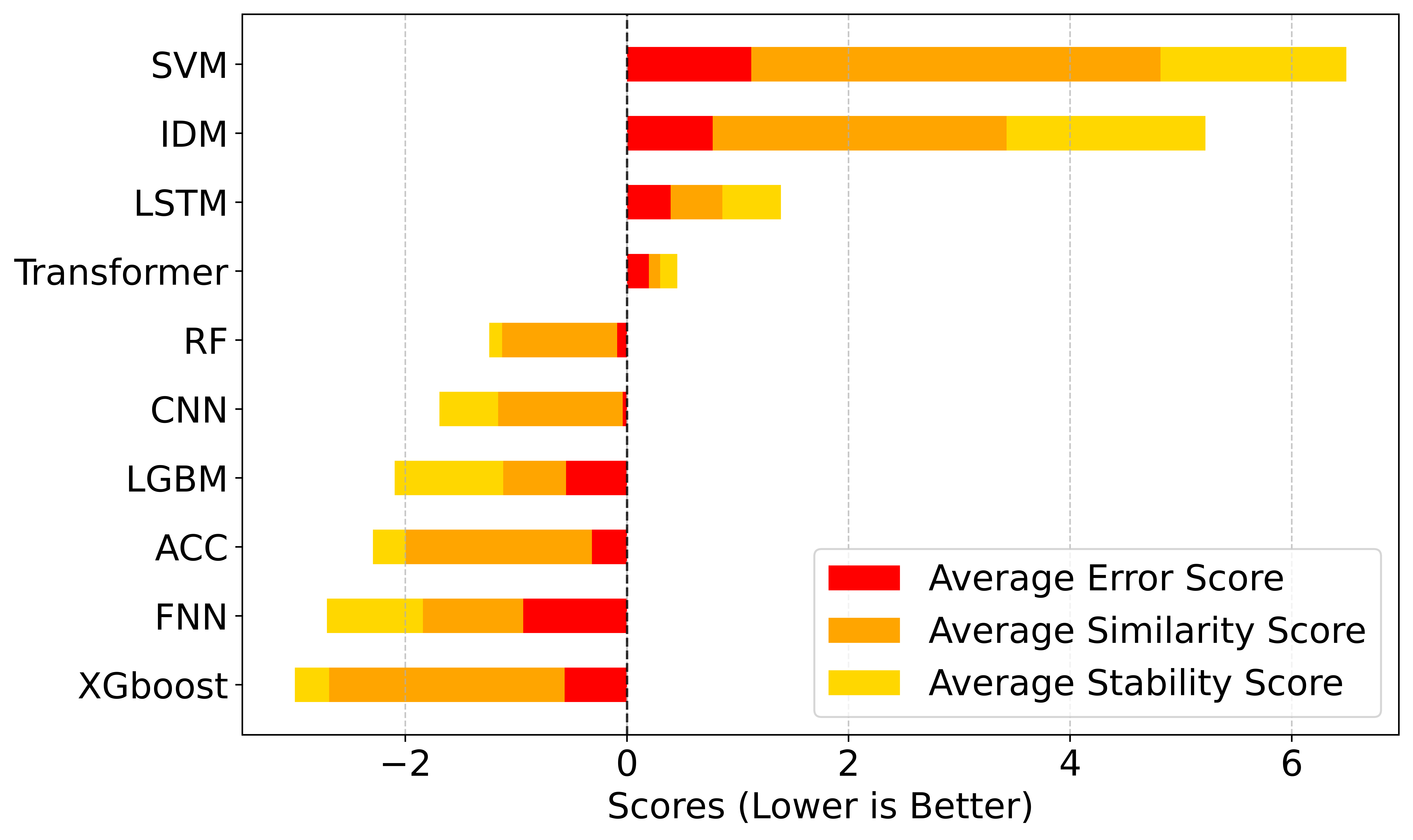}
    \caption{Aggregated Z-Scores for all models. Negative scores indicate better performance.}
    \label{fig:final_score}
\end{figure}

To rank overall model performance, the Z-scores for acceleration, speed, and position were averaged, with the results shown in Figure~\ref{fig:final_score}. The vertical line at zero represents the standardized baseline, where bars to the left indicate better-than-average performance and bars to the right indicate worse-than-average performance.

Figure~\ref{fig:final_score} shows that XGBoost and FNN achieve the lowest overall Z-scores, indicating the best balance of accuracy, stability, and trajectory realism. SVM and IDM perform worst due to high errors and instability, while LSTM excels in position similarity but suffers from acceleration instability. Transformer delivers balanced performance across metrics without leading in any single dimension. Overall, ML-based models outperform traditional car-following models, with tree-based methods excelling in acceleration prediction and neural networks providing better speed and position representation.
Table~\ref{tab:summary_zscore} summarizes the strengths and weaknesses of each model in replicating AS car-following trajectories.
\begin{table}
\caption{Summary of Comparative Model Performance Based on Z-Score Analysis}
\label{tab:summary_zscore}
\centering
\small
\renewcommand{\arraystretch}{1.25}
\begin{tabular}{p{2.3cm} p{4.7cm} p{7.0cm}}
\hline\hline
\multicolumn{1}{c}{Model} &
\multicolumn{1}{c}{Strengths} &
\multicolumn{1}{c}{Weaknesses} \\
\multicolumn{1}{c}{(1)} &
\multicolumn{1}{c}{(2)} &
\multicolumn{1}{c}{(3)} \\
\hline
XGBoost & Low error scores; strong overall stability. & Moderate trajectory-shape similarity in some dimensions. \\
FNN & Low errors across all dimensions; high trajectory similarity; very stable outputs. & Slightly higher variability in long-term position. \\
ACC & Simple structure; moderate accuracy and stability. & Less smooth and less accurate than learning-based models. \\
LGBM & Balanced performance in error, similarity, and stability; good accuracy. & High positional variability despite low error. \\
CNN & Good trend alignment in position with low positional error. & Poor acceleration stability and weak similarity. \\
RF & Moderate overall performance; more stable than SVM or IDM. & Noisy acceleration response; low trajectory similarity to real data. \\
Transformer & Strong trend-following in acceleration; stable and accurate acceleration prediction. & Slight instability and reduced precision in position. \\
LSTM & Excellent positional accuracy and shape similarity. & Poor acceleration prediction; high instability. \\
IDM & Simple model with low acceleration error. & Extremely unstable behavior; poor trend alignment. \\
SVM & Low average error in acceleration. & High errors and instability in speed and position; poor similarity. \\
\hline\hline
\end{tabular}
\normalsize
\end{table}

\subsection{Interpretation of Model Strengths and Weaknesses} \label{sec:model_discussion}

The applied multi-criteria framework enabled a systematic and comprehensive comparison of model performance. Results demonstrated that while certain models excelled in short-term acceleration prediction, others were more effective in reproducing trajectory smoothness and positional consistency.

 Tree-based ensemble methods, in general, achieved strong accuracy in acceleration prediction because they partition the feature space into fine-grained regions and capture nonlinear interactions without assuming a predefined functional form. In particular, XGBoost’s boosting framework iteratively corrects residual errors, allowing it to detect subtle nonlinearities between spacing, relative speed, and follower states \cite{Zhao2022XGBoost-DNNHighway,Ke2017LightGBM:Tree}. This explains its consistently superior performance across prediction error and stability metrics. However, their discrete structure limits temporal continuity, leading to weaker trajectory similarity

Neural network–based models exhibited complementary strengths. FNN achieved strong short-term accuracy by modeling complex nonlinear relationships, while CNN improved position prediction through local temporal feature extraction but showed instability in acceleration \cite{Qin2023AAbility}. LSTM effectively captured long-term dependencies, resulting in stable and accurate position estimates, though at the expense of responsiveness to short-term acceleration changes \cite{Tang2024ApplicationData}. Transformer models provided balanced accuracy and similarity by capturing global patterns; however, their higher complexity and data requirements limited performance gains under data sparsity and noise \cite{Rowan2025ASimulations}.

SVM produced lower accuracy, largely due to its reliance on margin maximization in a high-dimensional kernel space. The large number of support vectors obtained indicated poor separability and high dependence on noisy samples, which limited generalization and caused oversmoothing in acceleration variability \cite{Xue2019RapidData}. 
These issues may be attributed to overfitting to outliers, underfitting due to suboptimal kernel selection or hyperparameter tuning via the Optuna optimizer, or inherent limitations of SVM in modeling temporal patterns present in trajectory data.

Traditional models (IDM, ACC), in general, performed less effectively, indicating limitations of models calibrated primarily for human drivers or other AV types. While ACC captured some trajectory trends better than IDM, both models rely on simplified, quasi-linear assumptions about spacing and speed error that cannot replicate the operational variability of AS, resulting in systematic bias and instability when evaluated across speed and position.

Although the results indicated that XGBoost achieved the strongest overall performance for AS trajectories, it is important to note that no single model consistently outperforms others in all dimensions evaluated. Each class of models demonstrated strengths, weaknesses, and characteristic trade-offs shaped by its underlying architecture, supported in prior studies \cite{Lu2023LearningModel,Hart2024TowardsLearning,Rowan2025ASimulations}.

\section{Conclusions and Recommendations} \label{sec:conclusions}

Studies indicate that AS exhibit driving dynamics that differ significantly from both HDV and passenger AV \cite{Favero2025}. Developing adequately calibrated models for AS is essential to enable realistic simulation and to understand their impacts on traffic networks. However, existing research on AS is scarce, and no prior study has systematically evaluated ML performance on AS trajectories. 

To fill these gaps, this study (1) developed car-following models using real-world AS trajectory data, and (2) created a multi-criteria evaluation framework to assess models across multiple performance dimensions. The proposed multi-criteria framework established a transferable methodology that researchers can adopt to compare the performance of car-following models.

The proposed framework integrated metrics for error prediction, trajectory stability, and trajectory similarity. The results provided a more holistic assessment than traditional single-metric approaches. The main findings were:

\begin{itemize}
    
    \item Among the evaluated models, XGBoost achieved the strongest overall performance, balancing predictive accuracy, smoothness, and trajectory realism; FNN also performed well across most dimensions.
    
    \item Tree-based ensembles and FNNs effectively capture short-term acceleration dynamics but are more prone to cumulative errors over longer horizons.
    
    \item Sequential models (e.g., LSTM and Transformer) better capture temporal dependencies and positional stability, often at the expense of responsiveness to short-term acceleration changes.
    
    \item Traditional car-following models (IDM, ACC) and kernel-based methods (SVM) exhibited, in general, weaker accuracy and stability, highlighting the limitations of applying formulations developed for HDVs to AS.
    
\end{itemize}

This study has the following limitations: after data cleaning, the dataset represented a specific AS behavior during a limited data collection, which may constrain the ability to capture the full variability of AS behaviors; the models tested considered data from a specific operational environment, which may restrict their generalization to other traffic conditions, roadway settings, vehicle types, and environments.

Future work should validate these models on larger and more diverse datasets, consider different traffic conditions, and explore hybrid models. In addition, reinforcement learning–based models should be tested, as they may better capture adaptive decision-making in dynamic and uncertain environments. Finally, this study focused exclusively on longitudinal vehicle motion, as the AS under study did not perform lane changes. Future research could extend these models to account for lane-changing behavior.

\subsection{Data Availability Statement}

The dataset, models, and code that support the findings of this study are available in \url{https://github.com/renanfavero30/AS_car_following.git}.
The AS trajectory dataset used in this study was collected during a field deployment in Lake Nona, Orlando, Florida.

\subsection{Acknowledgments}

This research was supported by the University of Florida Transportation Institute (UFTI) and the Southeastern Transportation Research, Innovation, Development, and Education Center (STRIDE). The authors gratefully acknowledge the contributions of Beep in the Lake Nona deployment project for enabling access to the data used in this study.

\subsection{Disclaimer}

The contents of this paper reflect the views of the authors, who are responsible for the facts and the accuracy of the information presented herein. The statements, findings, conclusions, and recommendations are those of the authors and do not necessarily reflect the views or policies of the funding agencies or affiliated institutions.

\subsection{Author Contributions}

The authors confirm contributions to this paper as follows: study conception and design—Renan Favero and Lily Elefteriadou; data collection—Renan Favero; analysis and interpretation of results—Renan Favero and Lily Elefteriadou; manuscript preparation—Renan Favero and Lily Elefteriadou. All authors reviewed and approved the final version of the manuscript.

\subsection{Conflict of Interest}

The authors declare no potential conflicts of interest with respect to the research, authorship, and/or publication of this article.

\subsection{Funding}

This research was funded by the U.S. Department of Transportation (USDOT) under the Regional UTC STRIDE Project “Utilization of Connectivity and Automation in Support of Transportation Agencies’ Decision Making – Phase 2.”

%
%
\bibliography{references}

\end{document}